\documentclass[9pt,twocolumn,twoside]{pnas-new}

\templatetype{pnasmathematics} 

\setboolean{displaywatermark}{false}
\usepackage{physics}
\usepackage{amsmath}
\usepackage{caption}
\usepackage{subcaption}

\title{Two Hundred Years After Hamilton: The Simple Axiom That Underlies Classical Mechanics}

\author[a]{David J. Tannor}

\affil[a]{ Weizmann Institute of Science, Rehovot 76100, Israel}

\leadauthor{David J. Tannor}

\authorcontributions{}
\authordeclaration{The authors declare no conflict of interests.}
\correspondingauthor{\textsuperscript{1} E-mail: david.tannor@weizmann.ac.il}


\begin{abstract}
In 1834-1835, Hamilton published two papers that revolutionized classical mechanics. In these papers, he introduced the Hamilton-Jacobi equation, Hamilton’s equations of motion and the principle of least action. These three formulations of classical mechanics became the forerunners of quantum mechanics, but none of these is what Hamilton was looking for: he was looking for what he called the principal function, $S(q',q'',T)$, from which the entire trajectory history can be obtained just by differentiation. Here we show that all of Hamilton’s formulations can be derived just by assuming that the principal function is additive, $S(q',q'',T)=S(q',Q,t_1)+S(Q,q'',t_2)$ with $t_1+t_2=T$. This simple additivity axiom can be considered the fundamental principle of classical mechanics and shows that analytical mechanics is essentially just a footnote to the problem of finding the shortest path between two points. The simplicity of the formulation could provide new perspectives on some of the major themes in classical mechanics including symplectic geometry, periodic orbit theory and Morse theory, as well as giving new perspectives on quantum mechanics. Moreover, it could potentially provide a unified description of different areas of physics, leading to insight for example, into the transition from deterministic dynamics to statistical mechanics.
\end{abstract}

\dates{This manuscript was compiled on \today}

\begin{document}

\maketitle
\thispagestyle{firststyle}
\ifthenelse{\boolean{shortarticle}}{\ifthenelse{\boolean{singlecolumn}}{\abscontentformatted}{\abscontent}}{}

\section{Historical Introduction}
\label{sec:introduction}
\dropcap{H}amilton is probably best remembered for introducing what today is called the Hamiltonian, $H(q,p)$, leading to a beautifully symmetric reformulation of Newton's laws:
$\partial H/\partial p = dq/dt; \partial H/\partial q = -dp/dt.$  But historically, this was a parenthetical discovery in Hamilton's work. In two monumental papers published in 1834-35 \cite{ham1,ham2}, Hamilton introduced what were to become in the course of time three of the most important ways that we pass from classical mechanics to quantum mechanics: 1) Hamilton's equations of motion: when the numbers $q$ and $p$ are replaced by operators, these equations are the precursors of Heisenberg's equations of motion \cite{born-note}; 2) The Hamilton-Jacobi equation, a wave equation for matter and the precursor of the time-dependent Schr\"odinger equation \cite{schrodinger-note}; 3) The principle of least (or extremal) action, also known as Hamilton's principle \cite{nakane-note}, which is the precursor of the Feynman path integral formulation of quantum mechanics \cite{feynman-note} Yet none of these is what Hamilton was looking for. He was looking for a function he called the principal function, $S(q',q'',T)$, from  which the entire trajectory history can be obtained just by differentiation and elimination, with no integration \cite{lanczos-note}. In this article, we show that all three of the above formulations can be derived just by assuming that $S(q',q'',t)$ is additive, $S(q',q'',T) = S(q',Q,t_1)+ S(Q,q'',t_2)$ with $t_1+t_2=T$.
This simple additivity axiom can be considered as the fundamental principle of classical mechanics.
No further assumptions are necessary: momentum, energy, Hamiltonian, Lagrangian and action all emerge automatically. It is noteworthy that the equations of classical mechanics emerge in reverse order from conventional treatments. In particular, the principle of least (extremal) action appears at the end of our development, as opposed to conventional treatments where it is generally the starting point. It seems that analytical mechanics is essentially just a footnote to the most basic problem in the calculus of variations: that the shortest path between two points is a straight line.

\section{Thermodynamics: Entropy Maximum Principle}
\label{sec:entropy}
\begin{figure}[h!]
  \begin{center}
\includegraphics[width=16cm]{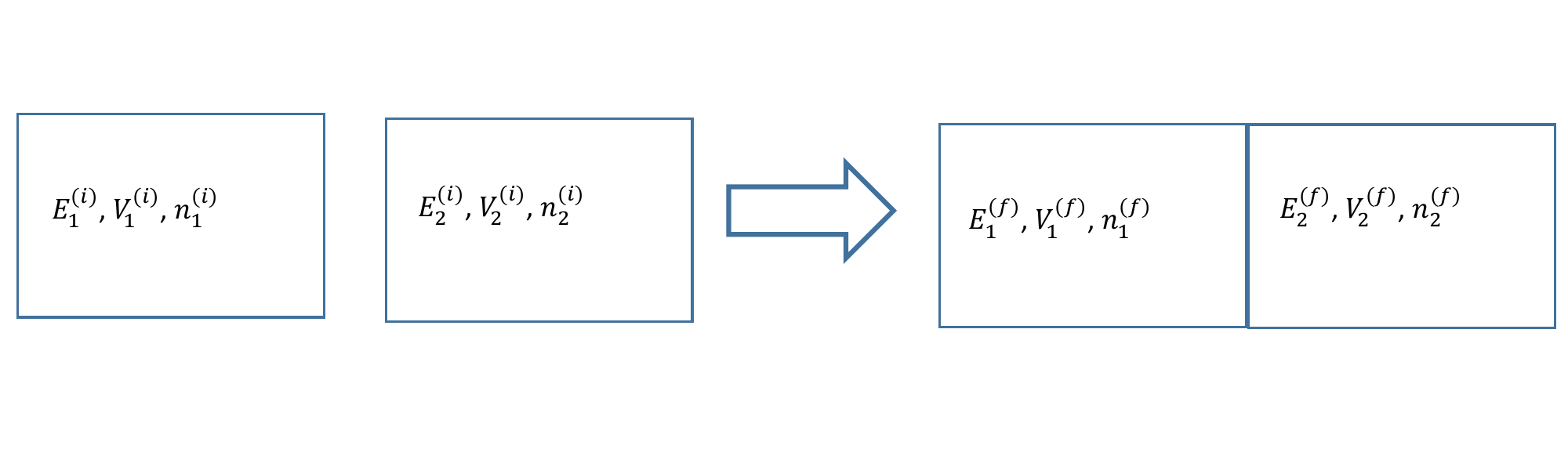}
\end{center}
\caption{\label{fig:composite_system} \footnotesize{Illustration of the entropy maximum principle in thermodynamics. On the left of the arrow is a composite system whose subsystems are not able to exchange $E$, $V$ or $n$. On the right, the same composite system after the two systems are brought into contact and allowed to exchange $E$, $V$ and $n$. (The superscripts $i$ and $f$ stand for initial and final, respectively). The entropy maximum principle answers the question, what will be the final values of $E$, $V$ and $n$ for each of the subsystems after they come into contact.}}
\label{fig1}
\end{figure}

\begin{figure}[h]
\begin{center}
\includegraphics[width=12cm]{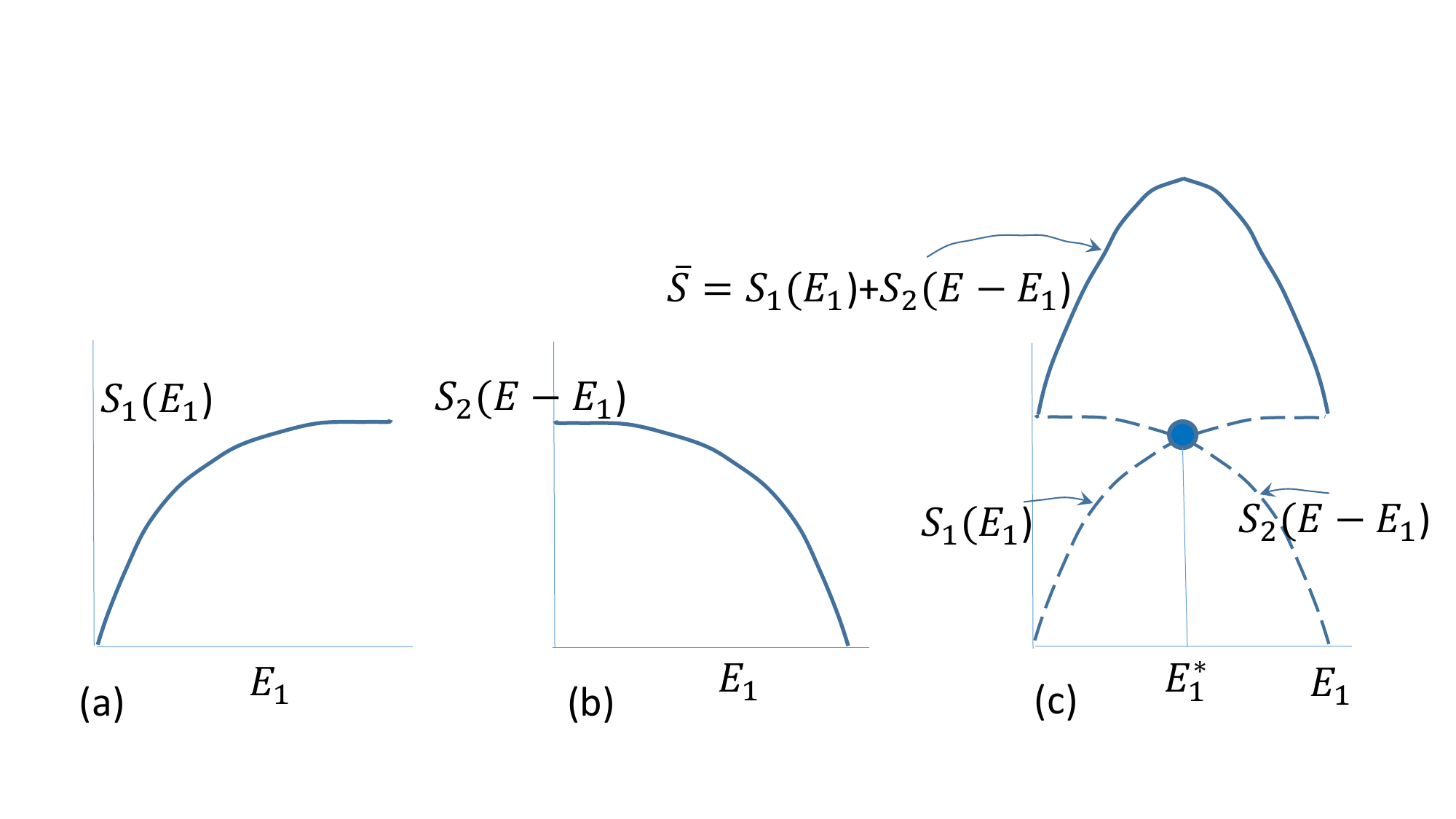}
\end{center}
\caption{\label{fig:S_min}Graphical illustration of the entropy maximum principle.
a)$S_1$ as a function of $E_1$; note that the function is concave. b) $S_2$ as a function of $E-E_1$. Since $E_1+E_2=E$, when $E_1$ gets larger $E-E_1=E_2$ gets smaller and therefore b) is the reflection of a). c) $S=S_1+S_2$ as a function of $E_1$. Because $S_1$ and $S_2$ are individually concave, their sum is concave and has an interior maximum at $E_1^*$, which determines the value of $E_2^*$; hence the entropy maximum principle determines the partitioning of the total energy $E$ between the two subsystems at equilibrium. }
\label{fig:entropy}
\end{figure}

Before proceeding to dynamics, it is convenient to introduce the methodology in the context of thermodynamics. Consider two subsystems brought into contact and allowed to exchange energy $E$, volume $V$ and material $n$ (see fig. \ref{fig:composite_system}). The entropy maximum principle answers the question: what will be the partitioning of $E$, $V$ and $n$ when equilibrium is reached \cite{callen,chandler}? For simplicity, assume that just $E$ can be exchanged between the subsystems. We posit that there exists a function $S(E)$ called the entropy that is additive for the two subsystems and that $S$ is a concave function of the energy, i.e. $S_1(E_1)$ and $S_2(E_2)$ are concave \cite{S-note}. The entropy maximum principle can then be expressed by the inequality,
\begin{equation}
S(E) \ge \bar{S} \equiv S_1(E_1)+S_2(E_2),
\end{equation}
i.e. $S(E)$ is the maximum of $S_1(E_1)+S_2(E_2)$ over all partitionings of the energy $E=E_1+E_2$ (see \ref{fig:entropy}). Equilibrium is characterized by the equality:
\begin{equation}
S(E) = S_1(E_1)+S_2(E_2),
\label{eq:se}
\end{equation}
i.e. at equilibrium, the entropy of the unconstrained composite system is equal to the sum of the entropies of the constrained subsystems.

To find the values of $E_1$ and $E_2$ at equilibrium, we maximize the combined entropy of the two subsystems, $\bar{S} = S_1(E_1)+S_2(E_2)$, subject to the constraint that the total energy is conserved, $E_1+E_2=E$. To this end, we define $\bar{\bar{S}}$:
\begin{equation}
\bar{\bar{S}}(E_1,E_2,\beta) = S_1(E_1)+S_2(E_2)-\beta(E_1+E_2-E),
\label{eq:h_thermo}
\end{equation}
where we have added the constraint equation with the Lagrange multiplier $\beta$.
Taking derivatives and setting them equal to zero:
\begin{equation}
(\partial \bar{\bar{S}}/\partial E_1)_{E_2,\beta}=0 = \partial S_1/\partial E_1-\beta,~~~~~~(\partial \bar{\bar{S}}/\partial E_2)_{E_1,\beta}=0 = \partial S_2/\partial E_2-\beta
\label{eq:e12*}
\end{equation}
\begin{equation}
(\partial \bar{\bar{S}}/\partial \beta)_{E_1,E_2} = E_1+E_2-E=0.
\label{eq:thermo_constraint}
\end{equation}
Because of the symmetry of $S_1(E_1)$ and $S_2(E_2)$ we may write eqs. \ref{eq:e12*} in a neutral form:
\begin{equation}
\partial S/\partial E-\beta = 0.
\label{eq:e*}
\end{equation}

Equations \ref{eq:e12*} together with eq. \ref{eq:thermo_constraint} provide three equations for three unknowns, $E_1$, $E_2$ and $\beta$.
Solving eqs. \ref{eq:e12*} for $E_1^*(\beta)$ and $E_2^*(\beta)$ and substituting into eq. \ref{eq:thermo_constraint} yields
$\beta^*$. Substituting $\beta^*$ along with $E_1^*(\beta^*)$, $E_2^*(\beta^*)$ into eq. \ref{eq:h_thermo} yields $\bar{\bar{S}}^*(E_1^*,E_2^*,\beta^*)=\bar{S}^*(E_1^*(\beta^*),E_2^*(\beta^*))$ where $\bar{S}^*(E_1^*(\beta^*),E_2^*(\beta^*))=S_1(E_1^*)+S_2(E_2^*)$
is the maximum of the constrained problem. By definition this is $S(E)$, hence we have \cite{note-quadratic}:
\begin{equation}
S(E)=S_1(E_1^*)+S_2(E_2^*).
\label{eq:se_star}
\end{equation}

The quantity $\beta$ has the physical interpretation of inverse temperature. To see this, note that at equilibrium the partitioning of energy $E_1^*$, $E_2^*$ between the two subsystems is that which maximizes the entropy. But from eqs. \ref{eq:e12*}, this is just the condition $\partial S_1/\partial E_1 = \partial S_2/\partial E_2$ --- that at equilibrium the inverse temperatures (and hence the temperatures) of the two subsystems are equal. Graphically, this is reflected in the equal and opposite slopes of $S_1(E_1)$ and $S_2(E-E_1)$ at $E_1^*$ in Fig. \ref{fig:entropy}c.

\section{Dynamics: Minimizing Hamilton's Principal Function}
\label{sec:action}
We now turn our attention to dynamics, and take as our starting point Hamilton's principal function (HPF). Inspired by the entropy maximum principle that requires only that the entropy be additive and concave, we assume only that HPF be additive and convex. Later on, we will even remove the assumption of convexity. As we shall see, all the equations of analytical mechanics will emerge, including that the action is the time integral of the Lagrangian.

\subsection{Minimizing HPF over partitionings of time}
\label{sec:action_lagrange}
\begin{figure}[h!]
  \begin{center}
\includegraphics[width=12cm]{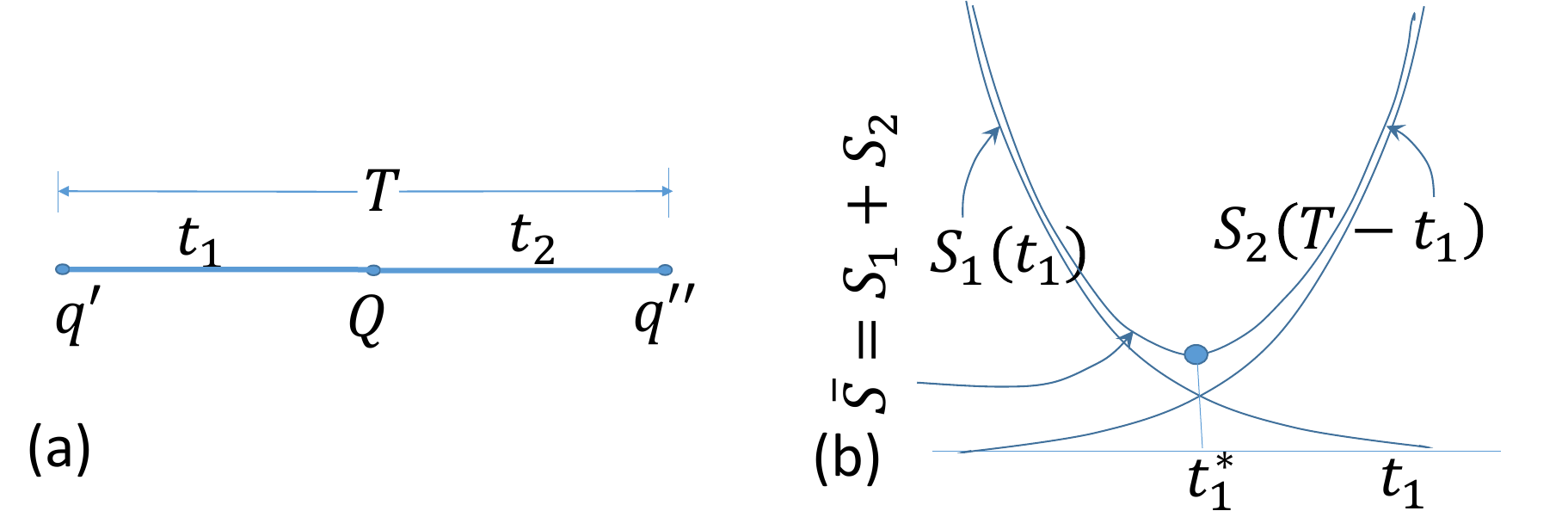}
\end{center}
\caption{a. Formulation of the HPF minimum principle. The endpoints $q'$, $q''$ and total time $T$ are fixed. An intermediate point $Q$ is fixed and we consider all possible partitionings of the total time $T$ such that $t_1+t_2=T$, where $t_1$ is the time from $q'$ to $Q$ and $t_2$ is the time from $Q$ to $q''$. b. With each of the two segments in a. there are associated HPFs, $S_1(q',Q,t_1)$ and $S_2(Q, q'', t_2)$ respectively, with the assumption that $S$ is convex with respect to all its arguments and that the HPF on each segment is minimized. The principle of minimal HPF can then be expressed as
$S(q',q'',T) \le S_1(q',Q,t_1) + S_2(Q,q'',t_2)$,
i.e. the composite action $S(q',q'',T)$ minimizes the total HPF over all partitionings of $t_1+t_2=T$.}
\label{fig:partitioning_t}
\end{figure}

Consider a convex function $S(q',q'',T)$ where $q'$ and $q''$ are $N$-dimensional vectors for a system of $N$ degrees of freedom and $T$ is the time to go from $q'$ to $q''$. By analogy with the entropy maximum principle, we partition the path into two segments, $[q',Q]$ and $[Q,q'']$, where again $Q$ is an $N$-dimensional vector. For the time being, we fix $Q$ and consider all possible partitionings of time between the two segments subject to the constraint that $t_1+t_2=T$.   We associate with each section its own HPF: $S_1(q',Q,t_1)$ and $S_2(Q, q'', t_2)$ (see Fig. \ref{fig:partitioning_t}). Note that in general $S_1$ and $S_2$ will be different functions. We will use a 1-dimensional notation for simplicity. In most cases, the explicitly multidimensional expressions are obvious, and at the end of this section we summarize the modifications necessary for the multidimensional expressions.


The principle of least HPF can then be expressed as:
\begin{equation}
S(q',q'',T) \le S_1(q',Q,t_1) + S_2(Q,q'',t_2).
\label{eq:hpf_inequality}
\end{equation}
To see why this is, note that in the true solution of the problem there is an optimum partitioning of time $T$ into $t_1^*+t_2^*$ so as to minimize $S$ for the entire length. Thus, any other partitioning of $T$ into $t_1+t_2$ will give a higher value for $S$ \cite{simplicity-note}.  For the actual partitioning of $T$ we obtain the equality:
\begin{equation}
S(q',q'',T) = S_1(q',Q,t_1) + S_2(Q,q'',t_2),
\label{eq:hpf_equality}
\end{equation}
which we call the additivity axiom. It states that the condition for an actual path is that the sum of the HPFs for two subsegments is equal to HPF for the full path \cite{path-note}.

To find the partitioning of $T$ into $t_1+t_2$ for which equality holds, we extremize $\bar{S}= S_1(q',Q,t_1) + S_2(Q,q'',t_2)$ in the presence of the constraint $t_1+t_2=T$:
\begin{equation}
\bar{\bar{S}} = S_1(q', Q, t_1) + S_2(Q,q'',t_2) + E(t_1+t_2-T),
\label{eq:bar_h}
\end{equation}
where $E$ is a Lagrange multiplier for the constraint on total time $T$.
Taking derivatives with respect to $t_1$, $t_2$ and $E$ and setting them equal to zero we obtain:
\begin{equation}
\partial \bar{\bar{S}}/\partial t_1 = \partial S_1/\partial t_1 +E=0,~~~~~~\partial \bar{\bar{S}}/\partial t_2 = \partial S_2/\partial t_2 +E=0
\label{eq:phj1}
\end{equation}
\begin{equation}
\partial \bar{\bar{S}}/\partial E= t_1+t_2-T=0.
\label{eq:phj3}
\end{equation}
We may write eqs. \ref{eq:phj1} in the neutral form:
\begin{equation}
\partial S/\partial t +E=0.
\label{eq:phj}
\end{equation}
Equation \ref{eq:phj} is well-known in classical mechanics with $E$ identified as the energy \cite{landau,gutzwiller,heller} and is a precursor of the Hamilton-Jacobi equation (to be obtained below in Section \ref{sec:hj}). Equation \ref{eq:phj3} is of course just the constraint equation. Equations \ref{eq:phj1} also deserve comment: note that the same value of $E$ appears in both equations. By analogy with thermodynamics, where we have seen that the maximum entropy can be expressed as the equality of the inverse temperature of the subsystems, the minimization of HPF can be expressed as the equality of the energy of the segments of the trajectory where energy is $-\partial S/\partial t$. In a separate publication, we will show how the method can be extended to time-dependent Hamiltonians.

Note that in passing from eq. \ref{eq:hpf_inequality} to eq. \ref{eq:hpf_equality} the inequality has been replaced by an equality. Equation \ref{eq:hpf_equality} is more general than eq. \ref{eq:hpf_inequality} in that it includes cases where HPF is an extremum, not a minimum. This corresponds to the result of Morse theory  \cite{morse,gutzwiller} that although $S(q',q'',T)$ is a minimum for short times, after reaching a caustic the minimum becomes a generalized saddle point. The terminology in the literature reflects this, where the principle of least action is often qualified as the principle of extremal or stationary action. Henceforth we will speak of extremizing HPF unless otherwise specified.

\subsection{Extremizing HPF over Partitionings of Intermediate Position $Q$}
\label{sec:action_q}

\begin{figure}[h!]
  \begin{center}
\includegraphics[width=10cm]{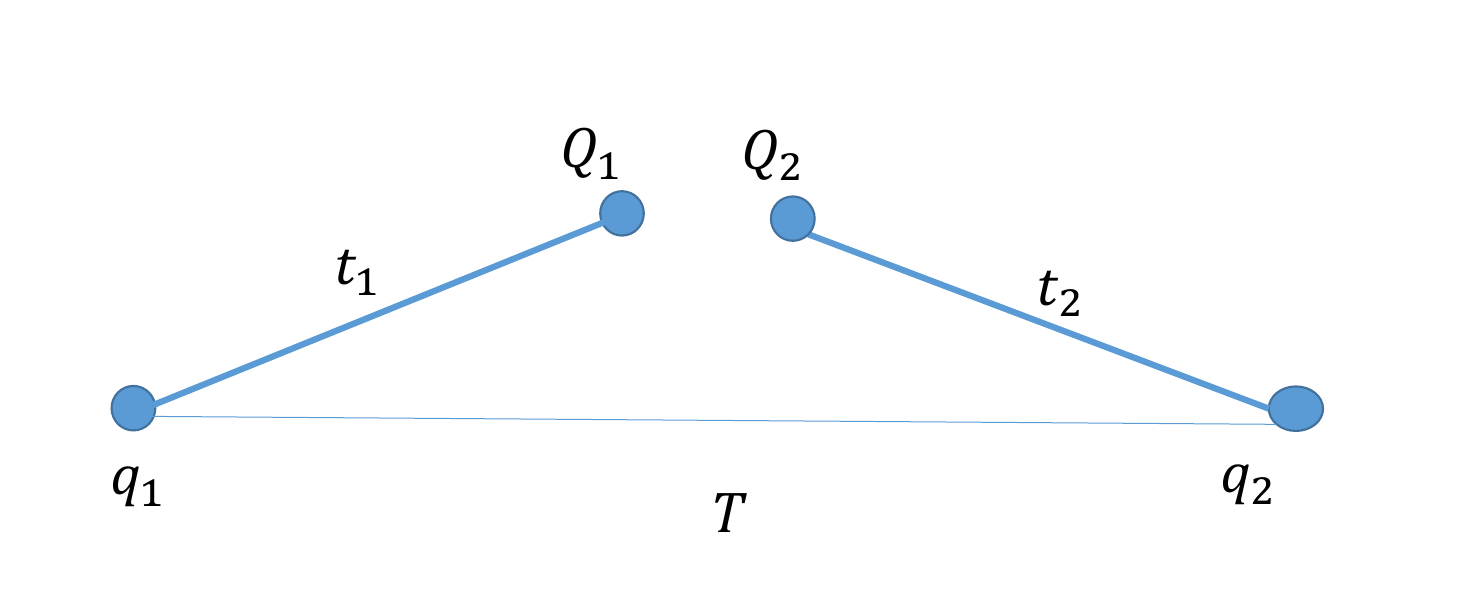}
\end{center}
\caption
{Formulation of the principle of extremal HPF, stage 2. $q'$, $q''$ and total time $T$ are fixed, but the endpoints $q_1$ and $q_2$ and the intermediate point $Q$ are now deconstrained. We consider all possible partitionings of the segment $[q_1,q_2]$ into $[q_1,Q_1]$ and $[Q_2,q_2]$, where $Q_1$ and $Q_2$ need not be identical (see text). With each of the two segments we associate HPFs, $S_1(q_1,Q_1,t_1)$ and $S_2(Q_2, q_2, t_2)$ respectively. For an actual trajectory, we then require $S(q_1,q_2,T) = S_1(q_1,Q_1,t_1) + S_2(Q_2,q_2,t_2),$
i.e. additivity of the HPFs with the constraints $Q_1=Q_2$ and $t_1+t_2=T.$}
\label{fig:line_segments_abc}
\end{figure}

We now fix the partitioning of time $T=t_1+t_2$ and allow the endpoints and the intermediate point to vary. Henceforth, we will use the notation $q_1$ and $q_2$ for the varying endpoints to distinguish them from the fixed endpoints $q'$ and $q''$. We consider several possible levels of allowing the intermediate point $Q$ to vary. The most straightforward is to allow $Q$ to vary but to constrain it to lie along the curve that extremizes $S(q_1,q_2,T)$. A second possibility, one level less constrained, is to allow $Q$ to lie off the curve that extremizes $S(q_1,q_2,T)$ \cite{geo-note}. Finally, we may deconstrain $Q$ even further, by not forcing endpoint $Q_1$ associated with $q_1$ to be equal to endpoint $Q_2$ associated with $q_2$ (see Fig. \ref{fig:line_segments_abc}). This seems to defy logic, since we must have a continuous curve from $q_1$ to $q_2$, but the apparent inconsistency that $Q_1 \neq Q_2$ presents no problem through the judicious use of Lagrange multipliers, and in fact contains the essential idea behind the general theory of canonical transformations \cite{treatment-note}.

Consider again the sum of the HPFs $\bar{S}$,
\begin{equation}
\bar{S}=S_1(q_1,Q_1,t_1) + S_2(Q_2,q_2,t_2)
\end{equation}
but now we specify the coordinate constraints explicitly:
\begin{equation}
q_1=q’, \hspace{1cm} q_2=q'', \hspace{1cm} Q_2=Q_1.
\end{equation}
We introduce the $N$-dimensional Lagrange multipliers $p_1$, $p_2$ and $P$:
\begin{equation}
\bar{\bar{S}}=S_1(q_1,Q_1,t_1) + S_2(Q_2,q_2,t_2) +p_1(q_1-q’) - p_2(q_2-q'')+P(Q_2-Q_1).
\end{equation}
Calculating the derivatives of $\bar{\bar{S}}$ with respect to $q_1$,$q_2$, $Q_1$ and $Q_2$ we obtain:
\begin{equation}
\partial \bar{\bar{S}}/\partial q_1 =\partial S_1/\partial q_1 + p_1 =0;~~~~~~~\partial \bar{\bar{S}}/\partial Q_1 = \partial S_1/\partial Q_1 - P=0
\label{eq:f1}
\end{equation}
\begin{equation}
\partial \bar{\bar{S}}/\partial q_2 =\partial S_2/\partial q_2 - p_2 =0;~~~~~~~\partial \bar{\bar{S}}/\partial Q_2 = \partial S_2/\partial Q_2 +P=0.
\label{eq:f1_a}
\end{equation}
Equations \ref{eq:f1} and eqs. \ref{eq:f1_a} are very similar, differing just in the sign of $p$ and $P$. We focus on eqs. \ref{eq:f1_a} since the subscript 2 corresponds to the final values of $q$ and $p$, in terms of which dynamical problems are generally formulated \cite{fraser-note-zero}.
Suppressing the subscripts we have:
\begin{equation}
p=\partial S/\partial q,~~~~~~P=-\partial S/\partial Q.
\label{eq:f1a}
\end{equation}
These equations are recognized as the standard equations for the first of four classical generating function (generally referred to as $F_1$) \cite{landau,goldstein,miller,heller}. Note that in addition we get the “handoff condition” \cite{heller},
\begin{equation}
P=-\partial S_2/\partial Q_2 = \partial S_1/\partial Q_1.
\end{equation}

\subsection{Inverting the Functional Forms: The Hamilton-Jacobi Equation}
\label{sec:hj}
We now proceed to derive the Hamilton-Jacobi equation. We first note that
the functional dependence $S=S(Q,q,t)$ implies that:
\begin{equation}
E=-\partial S/\partial t=E(Q,q,t),~~~~~~~~p=\partial S/\partial q=p(Q,q,t).
\label{eq:functional_e}
\end{equation}
Assume that the expression for $p(Q,q,t)$ can be inverted to give $Q=Q(p,q,t)$ \cite{conditions-note}; in $N$-dimensions this is an $N$-dimensional inversion. Substituting into $E(Q,q,t)$ gives
\begin{equation}
E(Q,q,t) = E(Q(q,p,t),q,t) \equiv H(p,q,t)=H(p(Q,q,t),q,t),
\label{eq:e=h1}
\end{equation}
where we have defined $H(p,q,t)\equiv E(Q,q,t)$. However, since $\partial S/\partial q = p$, it follows that
\begin{equation}
H(p,q,t) = H(\partial S/\partial q,q,t).
\end{equation}
Substituting into eq. \ref{eq:phj1} gives:
\begin{equation}
\partial S/\partial t + H(\partial S/\partial q,q, t)=0,
\label{eq:hj}
\end{equation}
which is the Hamilton-Jacobi equation \cite{landau,goldstein,heller}. Note that this is a PDE for $S$ with arguments $S(Q,q,t)$ \cite{fraser-note}..


\subsection{Deriving Hamilton’s Equations of Motion}
\label{sec:hamilton}
We now proceed to derive Hamilton’s equations of motion \cite{landau,goldstein,miller,heller}. Consider $S=S(q,Q,t)$ and note that
\begin{equation}
\partial \left( \partial S/\partial t \right) /\partial q = \partial \left( \partial S/\partial q \right)/ \partial t.
\label{eq:d2sa}
\end{equation}
This equating of cross-second derivatives taken in opposite orders is analogous to the procedure used to derive the Maxwell relations in thermodynamics \cite{callen,chandler}.
Using eqs. \ref{eq:phj}, \ref{eq:f1a} and \ref{eq:functional_e} and writing out the functional dependence explicitly we obtain
\begin{equation}
-(\partial E(Q,q,t)/\partial q)_{Q,t}= (\partial p(Q,q,t)/\partial t)_{Q,q},
\label{eq:maxwell}
\end{equation}
which looks very similar to $-\partial H/\partial q=dp/dt$. Hamilton's equations of motion do indeed emerge from eq. \ref{eq:maxwell}, but some additional care is needed. The key is to express on each of side of the equation one of the endpoints as a function of the other endpoint and its Lagrange multiplier: on the LHS, $Q=Q(p,q,t)$ and on the RHS, $q=q(P,Q,t)$. Again, these are in general $N$-dimensional inversions.

Consider first the LHS of eq. \ref{eq:maxwell}. Using the functional relationship in eq. \ref{eq:e=h1} we have:
\begin{equation}
-(\partial E/\partial q)_{Q,t}=
-(\partial H/\partial q)_{p,t} - (\partial H/\partial p)_{q,t}(\partial p/\partial q)_{Q,t}
\label{eq:lhs_a}
\end{equation}
Now consider the RHS of eq. \ref{eq:maxwell}. Viewing $q=q(P,Q,t)$ implies that $p(Q,q(P,Q,t),t)=p(Q,P,t)$. In Hamilton-Jacobi theory, the conjugate variables $P,Q$ are the initial conditions which indeed determine $p$ and $q$ at time $t$. Then
\begin{equation}
dp(Q,q(P,Q,t),t)/dt=(\partial p/\partial q)_{Q,t}(\partial q/\partial t)_{P,Q}+(\partial p/\partial t)_{q,Q},
\label{eq:dpdt}
\end{equation}
Motivated by the interpretation of $Q,P$ as initial conditions, we identify
\begin{equation}
(\partial q/\partial t)_{P,Q}=dq/dt.
\label{eq:dqdt}
\end{equation}
Substituting eqs. \ref{eq:lhs_a}, \ref{eq:dpdt} and \ref{eq:dqdt} into eq. \ref{eq:maxwell} we obtain:
\begin{equation}
-(\partial H/\partial q)_{p,t}-(\partial H/\partial p)_{q,t} (\partial p/\partial q)_{Q,t} =  dp/dt-dq/dt(\partial p/\partial q)_{Q,t}.
\label{eq:preham_1d}
\end{equation}
The factor $(\partial p/\partial q)_{Q,t}$ is common to both sides, so for the two sides to be equal:
\begin{equation}
-(\partial H/\partial q)_{p ,t}= dp/dt,~~~~~(\partial H/\partial p)_{q,t}=dq/dt.
\end{equation}
Somewhat unexpectedly, both of Hamilton’s equations of motion emerge from eq. \ref{eq:d2sa}. In fact, what emerges is an even more symmetrical form for Hamilton's equations of motion than is generally recognized:
\begin{equation}
-(\partial H(p,q,t)/\partial q)_{p ,t}= dp(P,Q,t)/dt,~~~~~(\partial H(p,q,t)/\partial p)_{q,t}=dq(P,Q,t)/dt,
\end{equation}
i.e. the LHS of both equations depends on the old variables, $p,q$ while the RHS depends on the new variables $P,Q$.  This idea of two old conjugate variables $p,q$ and two new conjugate variables $P,Q$ is generally encountered only in the theory of canonical transformations; it emerges that is actually explicit in Hamilton's equations of motion themselves \cite{baez-note}.

\subsection{The Lagrangian and the Explicit Form for the Action}
\label{sec:lagrangian}
We now return to $S=S(Q,q,t)$ and substitute the functional form $q=q(Q,P,t)$.  This gives $S(Q,q(Q,P,t),t)$. As a result, the total derivative of $S$ is given by:
\begin{equation}
dS/dt=(\partial S/\partial q)_{Q,t}(dq/dt)_{Q,P}+(\partial S/\partial t)_{q,Q}
=p \dot{q}-E = L,
\label{eq:lagrangian}
\end{equation}
where $L$ is the Lagrangian. Consequently,
\begin{equation}
S=\int_0^T (dS/dt) dt = \int_0^T (p \dot{q} - E) dt= \int_0^T L(q,\dot{q},t) dt,
\label{eq:s_int_l}
\end{equation}
which is the well-known formula expressing the action as the time integral of the Lagrangian \cite{landau,goldstein,heller}.

It is a short step from eq. \ref{eq:s_int_l}, $S=\int_0^T L(q,\dot{q},t)dt$, to the conventional formulation of the principle of least action (Hamilton’s principle). We derived eq. \ref{eq:s_int_l} based on the partitioning of $S(q_1,q_2,T)$ into two segments and the inequality $S(q_1,q_2,T) \le S_1(q_1,Q,t_1)+S_2(Q,q_2,t_2)$. Consider now dividing $S(q_1,q_2,T)$ into $N$ segments. The inequality then takes the form:
\begin{eqnarray}
S(q_1,q_2,T) & \le  S_A(q_1,q_A,t_A)+S_B(q_A,q_B,t_B)+S_C(q_B,q_C,t_C)+ \ldots + S_N(q_{(N)},q_{N+1},t_{N+1})  \nonumber \\
& =  \int_{0}^{t_A} L(q,\dot{q}) dt + \int_{t_A}^{t_B} L(q,\dot{q}) dt + \int_{t_B}^{t_C} L(q,\dot{q}) dt + \ldots + \int_{t_N}^{T} L(q,\dot{q}) dt.
\label{eq:least_action0}
\end{eqnarray}
Taking the limit $N \rightarrow \infty$ we obtain
\begin{equation}
S(q_1,q_2,T)= \min \int_{0}^{T} L(q,\dot{q}) dt
\label{eq:least_action}
\end{equation}
subject to the constraints $q_1=q'=q(t=0)$ and $q_2=q''=q(t=T)$. This is the conventional statement of the principle of least action. As discussed at the end of Section \ref{sec:action}\ref{sec:action_lagrange}, the more general formulation is to replace the inequality in eq. \ref{eq:least_action0} with an equality and the word mininum on the RHS of eq. \ref{eq:least_action} with the word extremal or stationary \cite{principle-note}.

We have come full circle. In many treatments of analytical mechanics, eq. \ref{eq:least_action} is the starting point \cite{landau,heller,goldstein}. We have arrived at this same formula, not as an axiom, but from an analysis which assumes just the additivity of $S(q_1,q_2,T)$, $S(q_1,q_2,T) = S(q_1,Q,t_1)+ S(Q,q_2,t_2)$ with $t_1+t_2=T$.

\section{Multidimensional Considerations}
As stated at the outset, for simplicity we adopted a 1-dimensional treatment but the extension to multidimensions in most cases is obvious. Here we elaborate on several points connected to the multidimensional formulation.

First, we show that the generating function relations eq. \ref{eq:f1a} are sufficient, within the assumption of invertibility, to completely determine the dynamics at all times. Writing the functional dependence explicitly:
\begin{equation}
p=\partial S(Q,q,t)/\partial q=p(Q,q,t);~~~~~~~P=\partial S/\partial Q=P(Q,q,t).
\end{equation}
Assuming that the second equation can be inverted (in $N$ dimensions this is an $N$-dimensional inversion) and adopting an explicitly multidimensional notation:
\begin{equation}
q_i=q_i(Q_1,\ldots,Q_N;P_1,\ldots,P_N;t).
\label{eq:q_i}
\end{equation}
Substituting into the first equation we obtain
\begin{equation}
p_i=p_i(Q_1,\ldots,Q_N;P_1,\ldots,P_N;t).
\label{eq:p_i}
\end{equation}
Equations \ref{eq:q_i} and \ref{eq:p_i}, taken together, amount to a complete solution of the dynamical problem: all the mechanical variables have been expressed as explicit functions of the time $t$ and the $2N$ initial conditions $Q_1,\ldots,Q_N;P_1,\ldots,P_N$ (or any $2N$ constants which can be adjusted to the initial conditions). This was Hamilton's guiding idea: to find a fundamental function, $S(Q,q,t)$, which can yield the values of the mechanical variables at all times by simple differentiations and eliminations, without any integration (\cite{lanczos-note}.

To derive Hamilton's equations of motion for the multidimensional case, we begin with the multidimensional generalization of eq. \ref{eq:preham_1d}:
\begin{equation}
-(\partial H/\partial q_i)_{\bar{\bf{q}},\bf{p},t}-\sum_{j=1}^N (\partial H/\partial p_j)_{\bf{q},\bar{\bf{p}},t} (\partial p_j/\partial q_i)_{\bar{\bf{q}},\bf{Q},t} =  dp_i/dt- \sum_{j=1}^N dq_j/dt (\partial p_i/\partial q_j)_{\bar{\bf{q}},\bf{Q},t},
\end{equation}
where the overbar signifies that $N-1$ variables are held fixed while one is derived. To obtain the $2N$ Hamilton equations of motion, we use the fact that
\begin{equation}
(\partial p_j/\partial q_i)_{\bar{\bf{q}},\bf{Q},t}=\partial^2 S(q_i,q_j)/\partial q_i \partial q_j = (\partial p_i/\partial q_j)_{\bar{\bf{q}},\bf{Q},t}.
\end{equation}
and then equate coefficients on both sides of the equations as in the 1-dimensional case.

Finally, the multidimensional generalization of eq. \ref{eq:lagrangian} is
\begin{equation}
dS/dt=\sum_{i=1}^N (\partial S/\partial q_i)_{\bar{\bf{q}},Q,t}(dq_i/dt)_{\bar{\bf{q}},Q,P}+(\partial S/\partial t)_{\bf{q},\bf{Q}}
=\sum_{i=1}^N p_i \dot{q}_i-E = L,
\end{equation}
giving the well-known multidimensional form for the Lagrangian.

\section{Conclusions}
\label{sec:conclusions}
We have shown that the Hamilton-Jacobi equation, generating functions for canonical transformations, Hamilton’s equations of motion, the Lagrangian and even the principle of least action (also known as Hamilton's principle) emerge from the additivity of Hamilton's principal functions, $S(q',q'',T)=S(q',Q,t_1)+S(Q,q'',t_2)$. We call the latter equation the additivity axiom, and consider it the single, fundamental axiom of analytical mechanics. Physics and mathematics have always valued deriving the maximum number of consequences from the simplest possible assumptions. For example, the five axioms of Euclid in geometry, the four Maxwell equations in electromagnetism, the three laws of Newton in classical mechanics, and the two postulates of Einstein in special relativity. The present formulation puts the subject of analytical mechanics on a similar footing with just a single, simple axiom.

It is intriguing to note that despite the collective historical perception of Hamilton's work, this was his central interest: to find a function $S(q',q'',T)$ that would allow the solution of the dynamical problem just by differentiations and substitutions \cite{lanczos-note}. In a very real sense, we have taken Hamilton’s program even further than Hamilton expected. There is no a priori assumption about velocity, momentum, energy, Hamiltonian or Lagrangian: these quantities emerge automatically. In fact, iterating eq. \ref{eq:hpf_equality} in principle provides a method for obtaining the trajectory motion from $S(q',q'',T)$ without even differentiating.

It is interesting to compare the treatment here with the problem of finding the shortest path between two points. Seen from a purely mathematical perspective, the latter can be formulated by introducing an intermediate point $Q$ and reexpressing the problem in terms of $q'$, $q''$ and $Q$. Since distance $d$ is additive, we may write $d(q',q'')=d(q',Q)+d(Q,q'')$ for $Q$ on the extremal path. Clearly, this equation can be generalized from a line to any geodesic. The structural similarity with the mechanical additivity axiom becomes even more pronounced if time is taken as the $n+1$ coordinate \cite{lanczos_time-note}. In this case, the mechanical additivity axiom takes the form $S(q',q'')=S(q',Q)+S(Q,q'')$, exactly the same as for a geodesic, with the physics entering only through the functional form of $S(q',q'')$ (which in turn determines $H(p,q)$ and $L(q, \dot{q})$). It would therefore appear that analytical mechanics is just a footnote to the mathematical problem of finding the shortest distance between two points. As such, this work should lead to a critical reevaluation of the place of analytical mechanics within the mathematics and physics literature.

One of the insights that emerges from the present treatment is that the structure of Hamilton’s equations arises from expressing one of the end points in $S(q',q'',T)$ (denoted $S(Q,q,t)$ in the paper) as a function of the other endpoint and its Lagrange multiplier. This gives rise to the triplets $(Q,P,t)$ and $(q,p,t)$. When Hamilton's equations of motion are represented using these triplets, an additional symmetry in the equations is revealed: one side is expressed in terms of $(q,p,t)$ and the other side in terms of $(Q,P,t)$. If this triplet structure is combined with a dual formulation of Lagrange multipliers \cite{hilbert-note}, i.e. if one or more of the primary variables $(q_1,q_2,t)$ is interchanged with its Lagrange multiplier $(p_1,p_2,E)$, we obtain the ``other half of classical mechanics''. For example, the four combinations of interchanging $(q_1,q_2)$ with $(p_1,p_2)$ give rise to the four generating functions for canonical transformations, $F_1-F_4$, \cite{landau,goldstein,miller,heller} while interchanging $t$ with $E$ gives the Mapertuis variational principle for the short action, $S(q_1,q_2,E)$ \cite{landau,goldstein,arnold}. A kind of Rubik's cube structure emerges to classical mechanics, as will be discussed in a future publication \cite{unsettled-note}.

The simplicity of the present formulation could potentially give a new perspective on some of the major themes in classical mechanics including symplectic geometry \cite{arnold,littlejohn}, periodic orbit theory \cite{gutzwiller,berry,smilansky,heller} and Morse theory \cite{morse,maslov}. By providing a different perspective on classical mechanics, this work could inspire alternative formulations of quantum mechanics with their own distinct advantages \cite{goldfarb,koch}. Moreover, because of the compact, geometrical character of the central equation, the approach could lead to a unified formulation of different areas of physics, ranging from mechanics to geometrical optics, and possibly electrodynamics and thermodynamics \cite{yourgrau_planck2-note}. For example, we have seen a series of striking analogies between the mathematics of classical mechanics and that of thermodynamics, with Hamilton’s principle function in dynamics playing the role of the entropy in thermodynamics, and convexity replacing concavity. In view of the sequential transition from convex to concave character every time a caustic is encountered, as described by Morse theory \cite{morse,gutzwiller}, a unified formulation could provide insight into the transition from deterministic dynamics to statistical mechanics.

We close with a quote from Max Planck, who laid exceptional emphasis upon the need for a methodical investigation into the scope and nature of the principle of least action \cite{yourgrau_planck1-note}: "As long as there exists a physical science, it has as its highest and most coveted aim the solution of the problem to condense all natural phenomena ... into one simple principle. Amid the more or less general laws which mark the achievements of physical sciences during the course of the last centuries, the principle of least action is perhaps that which, as regards form and content, may claim to come nearest to that ideal final aim of theoretical research" \cite{planck}.

\acknow{Financial support for this work came from the Israel Science Foundation (1094/16 and 1404/21), the German-Israeli Foundation for Scientific Research and Development (GIF) and the historic generosity of the Harold Perlman family. The author is grateful to Dr. Dahvyd Wing for helpful discussions throughout this work.}

\showacknow{} 

\bibliography{refs}

\begin{thebibliography}{10}
\bibitem{S-note}
$S_1$ and $S_2$ need not be the same function. If they are not we need to add
  an additional assumption, e.g. that the slopes of $S_1$ and $S_2$ vanish
  asymptotically at one limit.
\bibitem{note-quadratic}
The relationship between $E_1$,$E_2$ in eq. \ref{eq:se} and $E_1^*$,$E_2^*$ in
  eq. \ref{eq:se_star} is analogous to the relationship between $x$ in the
  equation $ax^2+bx+c=0$ and $x^*$ in the equation $x^*=(-b \pm \sqrt{b^2
  -4ac})/2a$, i.e. in the first equation $x$ is an unknown while in the second
  equation $x^*$ is the solution.
\bibitem{simplicity-note}
For simplicity, we may assume that $Q$ lies along the path from $q_1$ to $q_2$
  that minimizes HPF. If $Q$ does not lie along this path the argument is even
  stronger: in this case there is an optimal partitioning $T=t_1'+t_2'$ which
  in general is different from the optimal partitioning $t_1+t_2$ for $Q$ on
  the path; thus $t_1+t_2$ is a "frustrated" partition, compounding the
  suboptimality of the path.
\bibitem{path-note}
At this point, we are not assuming that the actual path is known, just that it
  exists and is characterized by eq. \ref{eq:hpf_equality}. This is completely
  analogous to our discussion of thermodynamics, where initially we do not
  assume that the equilibrium state is known, just that it exists and is
  characterized by eq. \ref{eq:se}.
\bibitem{treatment-note}
This treatment of $Q_1$ and $Q_2$ independently is perhaps also implicit in
  Jacobi's work. See footnote \cite{fraser-note}.
\bibitem{unsettled-note}
One might feel unsettled, that the recovery of the equations of classical
  mechanics in this paper depended on the choice of symbols used to represent
  the Lagrange multipliers. To this point we offer several responses. 1) In
  analytical mechanics, the momentum $p$ is normally introduced via the
  Legendre transform from $L(q,\dot{q})$ to $H(q,p)$, and hence is already an
  abstract mathematical construct. Moreover, in statistical mechanics the
  Lagrange multipliers used in the Gibbs entropy maximum principle turn out to
  be the intensive variables (e.g. $1/T$,$p/T$,$\mu/t$). So there is a good
  precedent for equating Lagrange multipliers (or conjugate Legendre variables)
  with physical quantities. From a functional point of view, if all the
  equations we have derived are isomorphic to the equations of classical
  mechanics, for all intents and purposes it is classical mechanics. Another
  potential criticism of this work is to say that the paper merely exchanges
  the time-tested Legendre transforms in analytical mechanics with Lagrange
  multipliers, but what has been gained? There are several compelling answers,
  but for brevity we give just one: since the Lagrange multipliers are each
  associated with a constraint, their number and character is determined a
  priori and they have a well-grounded physical basis. In contrast, when and
  why Legendre transforms are invoked in the conventional approach ($\dot{q}
  \rightarrow p$, $L \rightarrow H$, $S(t) \rightarrow S(E)$, $F_1 \rightarrow
  F_2-F_4$) is not clear since they are not tied to constraints.
\end{thebibliography}

\end{document}